%% file: main.tex
\documentclass[letterpaper]{article} 
\usepackage[]{aaai24}  
\usepackage{times}  
\usepackage{helvet}  
\usepackage{courier}  
\usepackage[hyphens]{url}  
\usepackage{graphicx} 
\usepackage{natbib}  
\usepackage{caption} 
\usepackage{algorithm}
\usepackage{algorithmic}
\usepackage{booktabs}       
\usepackage{amssymb}
\usepackage{amsfonts}       
\usepackage{nicefrac}       
\usepackage{microtype}      

\usepackage{newfloat}
\usepackage{listings}
\DeclareCaptionStyle{ruled}{labelfont=normalfont,labelsep=colon,strut=off} 
\floatstyle{ruled}
\newfloat{listing}{tb}{lst}{}
\floatname{listing}{Listing}
\input{macros}

\title{Extreme Solar Flare Prediction Using Residual Networks with HMI Magnetograms and Intensitygrams}

\def\authorEmail{sheen@korea.kr}

\author{
    Juyoung Yun\textsuperscript{\rm 1}, 
    Jungmin Shin\textsuperscript{\rm 2}\thanks{Corresponding author. E-Mail: \authorEmail}
}
\affiliations{
    \textsuperscript{\rm 1}Stony Brook University, New York, USA\\
    \textsuperscript{\rm 2}Defense Acquisition Program Administration (DAPA), Gwacheon-si, Republic of Korea
}

\usepackage{bibentry}

\usepackage{framed}

\begin{document}

\maketitle

\begin{abstract}
\input{Contents/0-abstract}
\end{abstract}

\section{Introduction}
\input{Contents/1-intro}

\section{Related Works}
\input{Contents/2-related}

\section{Methodology}

\input{Contents/3-method}

\section{Results}
\input{Contents/4-results}

\section{Discussion}
Our study highlights the potential of using HMI magnetograms and intensitygrams for the prediction of extreme solar flares. However, there are several limitations that need to be addressed in future research. One significant limitation is the overall scarcity of magnetogram training data, which constrains the robustness and generalizability of our model. To mitigate this, we propose the creation of a comprehensive solar image dataset specifically designed for research purposes. This dataset would provide a valuable resource for the scientific community, enabling more extensive and varied training of predictive models. Moreover, expanding our research with a richer magnetogram dataset could enhance the model's capability to capture the critical moments leading up to a flare. By focusing on the magnetogram data just before flare onset, we could develop models that not only predict the occurrence of solar flares but also provide more precise forecasts of the intense sunspot eruptions immediately preceding these events. 

\section{Conclusion}
\input{Contents/5-conclusion}

\section*{Acknowledgement}
We would like to thank Roman Bolzern and Michael Aerni from the Institute for Data Science, FHNW, Switzerland for providing the SDOBenchmark dataset. We also acknowledge the SDO satellite mission and JSOC Stanford for providing the raw data.


\bibliography{aaai24}

\end{document}

%% file: macros.tex
\usepackage{microtype}
\usepackage{graphicx}
\usepackage{subfigure}
\usepackage{booktabs} 
\usepackage[linewidth=1pt]{mdframed}

\usepackage{amsmath}
\usepackage{amssymb}
\usepackage{mathtools}
\usepackage{amsthm}

\usepackage[capitalize,noabbrev]{cleveref}

\theoremstyle{plain}

\theoremstyle{definition}

\theoremstyle{remark}

\usepackage[textsize=tiny]{todonotes}


%% file: Contents/0-abstract.tex
Solar flares, especially C, M, and X class, pose significant risks to satellite operations, communication systems, and power grids. We present a novel approach for predicting extreme solar flares using HMI intensitygrams and magnetograms. By detecting sunspots from intensitygrams and extracting magnetic field patches from magnetograms, we train a Residual Network (ResNet) to classify extreme class flares. Our model demonstrates high accuracy, offering a robust tool for predicting extreme solar flares and improving space weather forecasting. Additionally, we show that HMI magnetograms provide more useful data for deep learning compared to other SDO AIA images by better capturing features critical for predicting flare magnitudes. This study underscores the importance of identifying magnetic fields in solar flare prediction, marking a significant advancement in solar activity prediction with practical implications for mitigating space weather impacts.

%% file: Contents/1-intro.tex
Solar flares are intense bursts of radiation caused by the release of magnetic energy associated with sunspots. These events can significantly disrupt satellite communications, navigation systems, and power grids on Earth~\cite{noaan, Schwenn, Rama, Filjar1}. The severe consequences of solar flares, especially extreme C, M, and X class events, underscore the importance of accurate prediction to protect infrastructure and mitigate their impacts~\cite{c2023towards}. Additionally, solar flares pose risks to space weather, affecting technological systems, astronauts, and high-altitude flights~\cite{Schwenn, Rama}. Effective prediction allows for timely warnings and protective measures, such as adjusting satellite orbits and safeguarding power grids, thereby minimizing potential damage and ensuring the continued operation of critical services~\cite{Filjar1}.

The need for accurate and timely space weather forecasts is growing with our increasing reliance on satellite technology and high-altitude aviation. Solar activity, including flares and coronal mass ejections, can influence the Earth's magnetosphere and ionosphere, leading to geomagnetic storms that affect satellite operations, GPS accuracy, and ground-based technologies~\cite{Schwenn, Rama}. Recent advancements in solar observations from the Solar Dynamics Observatory (SDO) have provided high-resolution data, particularly from the Helioseismic and Magnetic Imager (HMI)~\cite{he2016deep}. HMI captures intensitygrams, detailing sunspots, and magnetograms, mapping the Sun's magnetic field. These datasets are crucial for understanding the magnetic complexity leading to solar flares~\cite{noaan, Schwenn, Rama}.

In this study, we propose a novel approach to predict extreme solar conditions, specifically C, M, and X class solar flares, by leveraging HMI intensitygrams (HMII) and magnetograms (HMIB). Our methodology involves detecting sunspots from intensitygrams and segmenting corresponding magnetic field patches from magnetograms. These patches are then used to train a Residual Network (ResNet)~\cite{he2016deep}, a deep learning model known for its effectiveness in image classification. We focus on a binary classification task by grouping Quiet, A, and B class events together and C, M, and X class events together to predict extreme flare occurrences. This approach emphasizes the importance of magnetic field analysis for accurate solar flare prediction.

%% file: Contents/2-related.tex
\begin{figure*}[t]
\centering
\includegraphics[width=1.8\columnwidth]{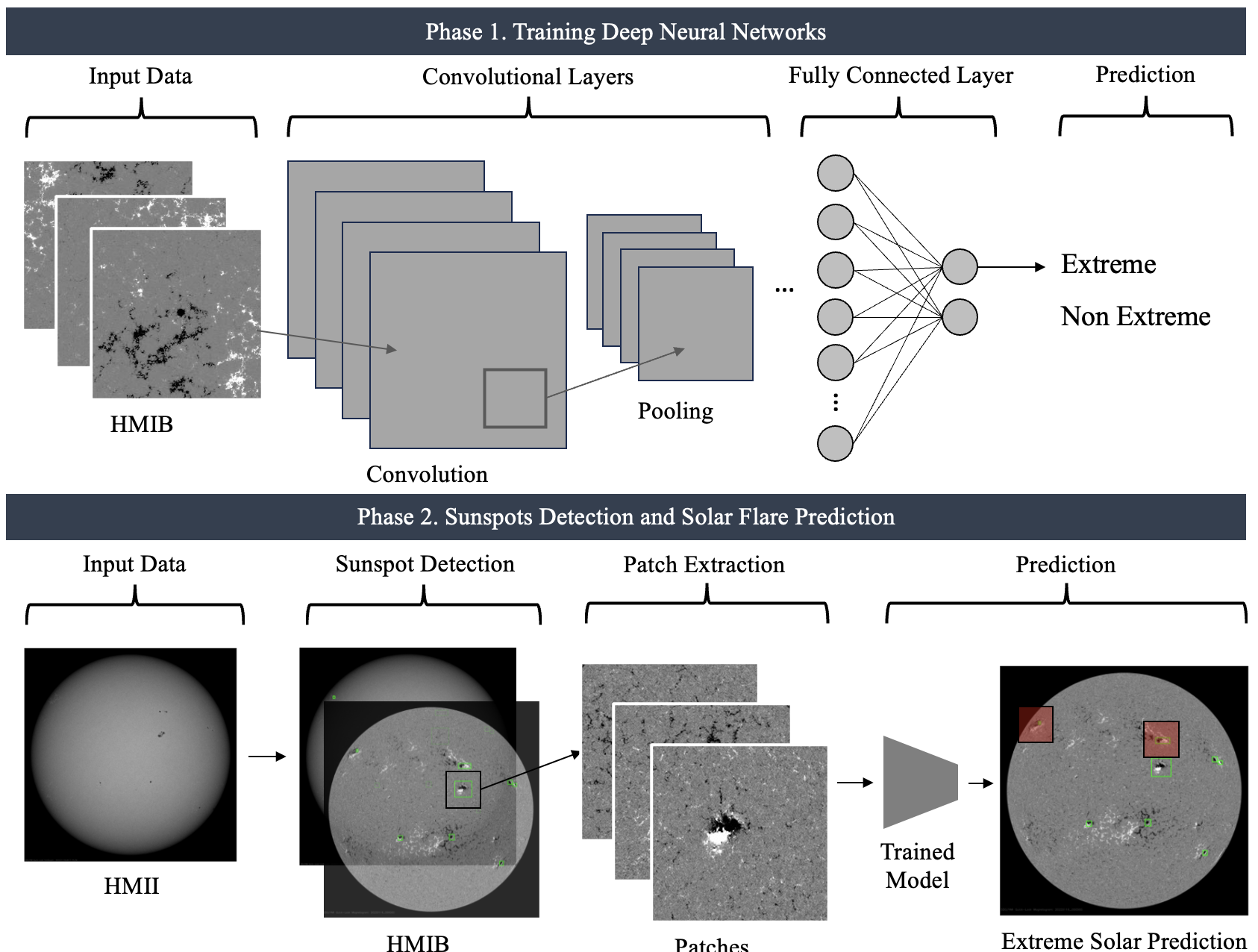}
\caption{The two-phase process for extreme solar flare prediction. Phase 1 involves training a deep neural network (ResNet) with HMI magnetogram (HMIB) data to classify extreme and non-extreme events. Phase 2 includes detecting sunspots on HMI intensitygram (HMII) images, extracting corresponding magnetic field patches from HMIB, and using the trained model to predict extreme solar flare occurrences.}
\label{fig:method}
\end{figure*}

In recent years, solar flare prediction has significantly advanced with the application of machine learning and deep learning techniques, leveraging data from the Solar Dynamics Observatory (SDO)~\cite{Nasa}, particularly the Helioseismic and Magnetic Imager (HMI). Various studies have explored the utility of HMI data for predicting solar flares.

Bobra and Couvidat used SHARP vector magnetic field data from HMI to calculate 25 physical parameters, including total flux and proxies for helicity and energy, training a machine learning algorithm to predict solar flares based on these parameters \cite{bobra2014sharps}. Nishizuka et al. developed the Deep Flare Net (DeFN), employing deep learning for operational solar flare prediction using 79 physics-based features extracted from HMI NRT data \cite{nishizuka2018operational}. Additionally, Li et al. focused on enhancing the interpretability of predictions using attention mechanisms within deep neural networks \cite{c2023towards}.

These approaches often rely on a broad range of features extracted from HMI data, combining both magnetograms and intensitygrams. Bobra and Couvidat’s method emphasizes physical parameters derived from SHARP data, while Nishizuka et al.'s DeFN integrates a large number of physics-based features. Li et al.'s approach is distinct in its use of attention mechanisms to improve interpretability, allowing for better understanding of the decision-making process within the neural network.

\begin{figure*}
\centering
\includegraphics[width=2.0\columnwidth]{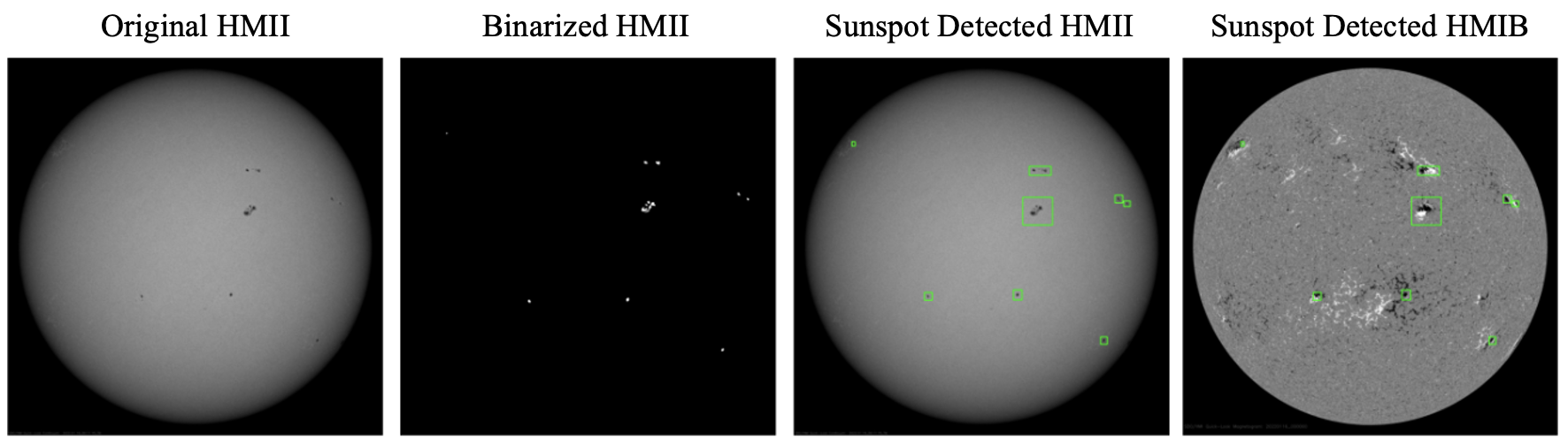}
\caption{The process of detecting sunspots using HMI intensitygrams. From left to right: the original HMI intensitygram (HMII), the binarized HMII, detected sunspots on the HMII, and the corresponding sunspot detection mapped onto the HMI magnetogram (HMIB).}
\label{fig:detect}
\end{figure*}

Our study is different because it specifically targets the prediction of extreme C, M, and X class solar flares using a more focused approach based on magnetic field patches. Unlike previous methods that use a wide range of image, we concentrate on detecting sunspots from intensitygrams and segmenting the corresponding magnetic field patches from magnetograms. These patches are then used to train and predict a Residual Network (ResNet)~\cite{he2016deep}, a deep learning model known for its effectiveness in image classification. By focusing on sunspot detection and magnetic field patch extraction, our method simplifies the feature selection process, potentially reducing the complexity and computational load compared to models that handle a broader set of features. 

%% file: Contents/3-method.tex
Our approach involves a two-step process as shown in Figure~\ref{fig:method}: training a Residual Network~\cite{he2016deep} on the SDO dataset~\cite{Nasa} to predict solar flare occurrences and using computer vision techniques to detect sunspots on HMI intensitygrams and segment magnetic field patches from HMI magnetograms. These patches are then used with the trained model to predict extreme flare occurrences.

\subsection{Training Deep Neural Networks with Magnetogram}
We utilized the ResNet50 architecture~\cite{he2016deep} for training on HMI magnetogram patch images to predict extreme solar flares. Magnetogram patch images were sourced from the Solar Dynamics Observatory (SDO) and resized to 256 × 256 pixels. The images were annotated with flare classifications based on their intensity levels, grouped into two categories: Quiet, A, and B class events as QA, and C, M, and X class events as MX. This binary classification aids in predicting extreme flare occurrences more effectively. The ResNet50 model, with weights pre-trained on the ImageNet dataset~\cite{deng2009imagenet}, was used. Final Fc layer consists of 512 neurons and ReLU activation, and an output layer is with a softmax activation function to classify images into QA or MX. 

\subsection{Sunspot Detection for Patch Extraction}
Our methodology for detecting sunspots using HMI intensitygrams involves key steps as shown in Figure~\ref{fig:detect}. We start with real-time HMI intensitygram images, convert them to RGB color space, and apply a noise-canceling filter~\cite{Buades}. The denoised images are then binarized using adaptive thresholding~\cite{Gonzalez} to distinguish sunspots from the background. Contours on the binary image~\cite{marr} detect sunspots, which are filtered by location within the solar disk. Close bounding rectangles are merged. Using detected sunspots, we extract 256x256 patches from HMI magnetogram images centered on the sunspot locations. These patches are fed into a trained ResNet to predict extreme solar flare occurrences.


%% file: Contents/4-results.tex
\subsection{Experimental Setting}
We used a ResNet-50 model~\cite{he2016deep} pretrained on the ImageNet dataset~\cite{deng2009imagenet} to train our solar flare prediction model. The training was conducted using the SDOBenchmark dataset~\cite{SDOBenchmark}, which consists of 256x256 pixel images. The model was trained for 20 epochs with a batch size of 32. The Adam optimizer~\cite{kingma2014adam} was employed with a learning rate of 0.0001. The experiments were performed on an Nvidia RTX 4080 GPU.

\begin{figure}
\centering
\includegraphics[width=1.0\columnwidth]{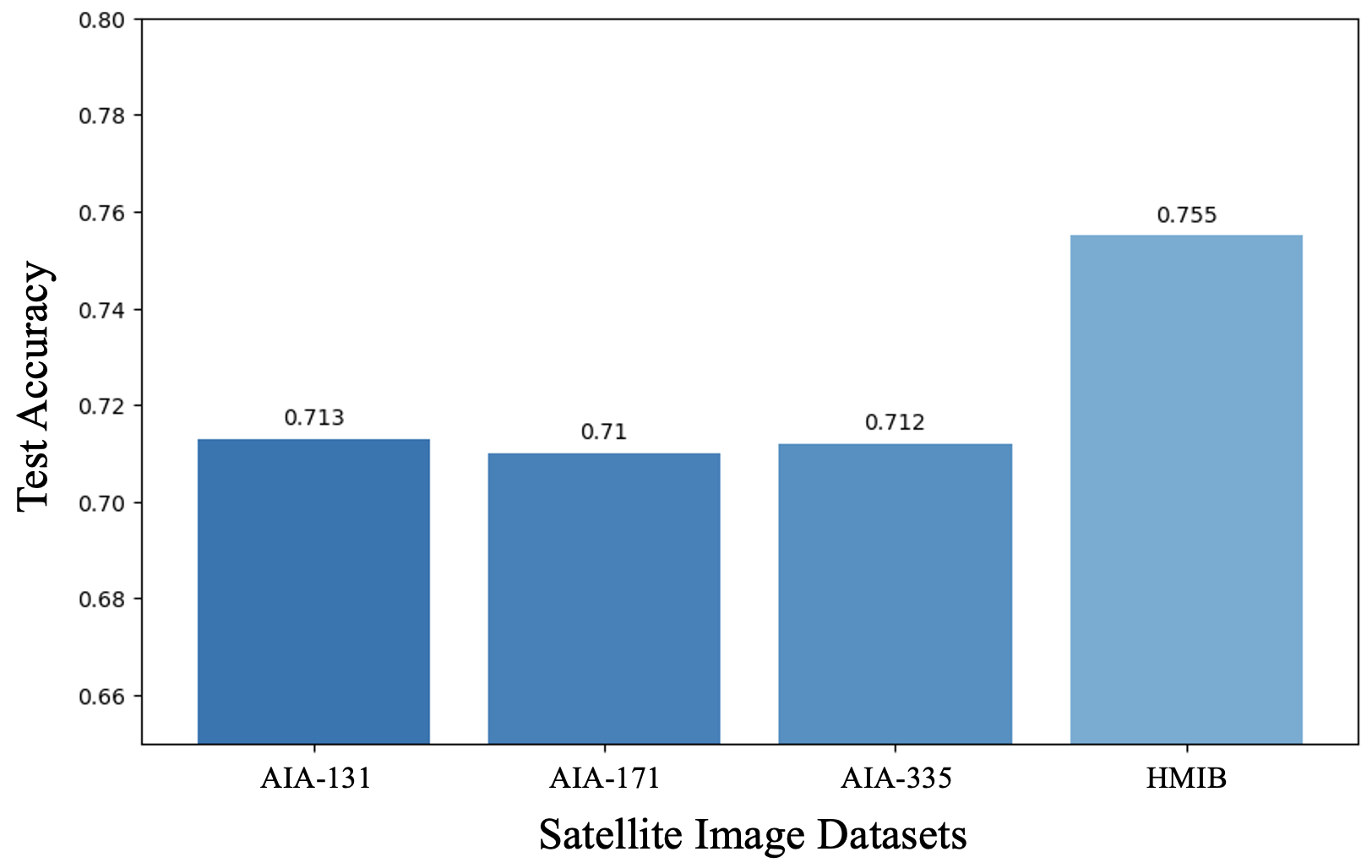}
\caption{Comparison of test accuracy for different satellite image datasets using the SDOBenchmark dataset~\cite{SDOBenchmark}. The model used is ResNet-50~\cite{he2016deep} pre-trained on ImageNet~\cite{deng2009imagenet} for predicting two flare classes, QA and MX.}
\label{fig:aia}
\end{figure}

\begin{figure*}
\centering
\includegraphics[width=2.0\columnwidth]{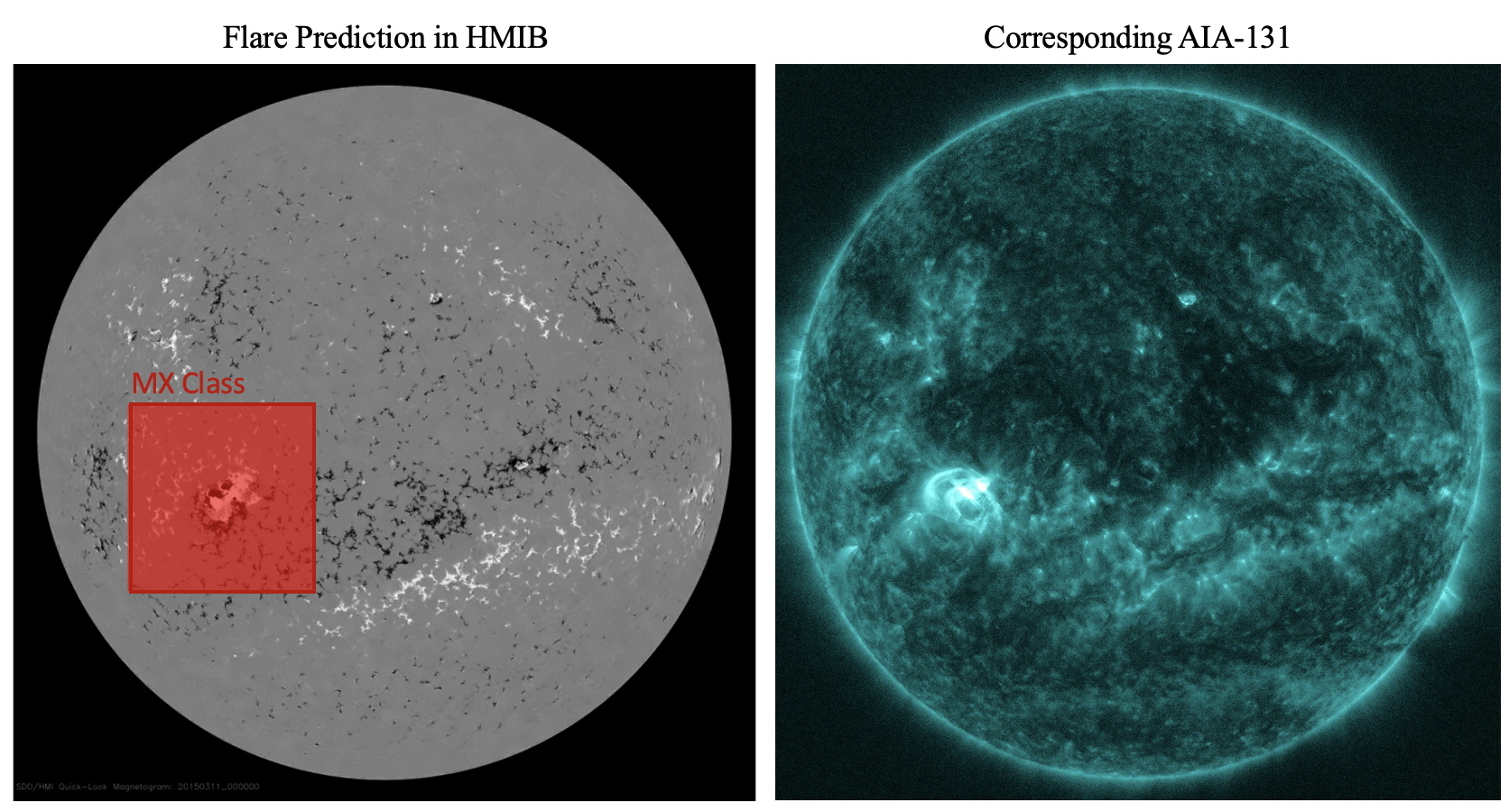}
\caption{Flare Prediction Results. (Left) HMI magnetogram (HMIB) showing predicted MX class flare region. (Right) Corresponding AIA-131 image showing the flare. Timeline is 2022-01-16 00:00:00.}
\label{fig:results}
\end{figure*}

\subsection{Training Results of Deep Neural Networks}
The training results of our deep neural networks demonstrate that using HMI magnetogram (HMIB) patches for predicting extreme solar flares yields higher accuracy compared to other satellite image datasets. As shown in Figure~\ref{fig:aia}, the test accuracy achieved with HMIB images is 0.755, significantly higher than the accuracies obtained with various AIA datasets, such as AIA-131 (0.713), AIA-171 (0.710), and AIA-335 (0.712). This underscores the importance of magnetic field data captured in HMIB for accurate solar flare prediction.

\subsection{Prediction Results}
Our model demonstrated ability in predicting MX (C, M, and X) class solar flares using HMI magnetograms and intensitygrams. Figure~\ref{fig:results} shows the final prediction result, with the left panel showing the HMI magnetogram (HMIB) where the predicted flare region is highlighted, and the right panel displaying the corresponding AIA-131 image that confirms the flare occurrence.

%% file: Contents/5-conclusion.tex
In this study, we proposed a novel approach for predicting extreme solar flares using HMI intensitygrams and magnetograms. By detecting sunspots from intensitygrams and extracting corresponding magnetic field patches from magnetograms, we trained a Residual Network (ResNet) to classify these flares with high accuracy. Our experimental results demonstrated that HMI magnetograms provide superior data for deep learning models compared to other SDO AIA images, due to their ability to capture critical magnetic field features necessary for accurate flare prediction. Our research underscores the importance of magnetic field data in forecasting solar flare magnitudes. This advancement in solar activity prediction has practical implications for mitigating the impacts of space weather.